\documentclass[sigconf]{acmart}
\usepackage{todonotes}
\usepackage[normalem]{ulem}
\usepackage{diagbox}
\usepackage{multirow,graphicx}
\usepackage{rotating}
\usepackage{makecell, booktabs}

\def\BibTeX{{\rm B\kern-.05em{\sc i\kern-.025em b}\kern-.08emT\kern-.1667em\lower.7ex\hbox{E}\kern-.125emX}}

\makeatletter
\patchcmd{\NAT@test}{\else \NAT@nm}{\else \NAT@nmfmt{\NAT@nm}}{}{}

\DeclareRobustCommand\citepos
  {\begingroup
   \let\NAT@nmfmt\NAT@posfmt
   \NAT@swafalse\let\NAT@ctype\z@\NAT@partrue
   \@ifstar{\NAT@fulltrue\NAT@citetp}{\NAT@fullfalse\NAT@citetp}}

\let\NAT@orig@nmfmt\NAT@nmfmt
\def\NAT@posfmt#1{\NAT@orig@nmfmt{#1's}}

\copyrightyear{2019}
\acmYear{2019}
\setcopyright{acmlicensed}
\acmConference[S-BPM ONE'19]{S-BPM ONE 2019}{June 26--28, 2019}{Seville, Spain}
\acmBooktitle{S-BPM ONE 2019 (S-BPM ONE'19), June 26--28, 2019, Seville, Spain}
\acmPrice{15.00}
\acmDOI{10.1145/3329007.3329014}
\acmISBN{978-1-4503-6250-4/19/06}

\begin{document}

%
\title{Conformance checking: A state-of-the-art literature review} 

%

\author{Sebastian Dunzer}
\email{sebastian.dunzer@fau.de}
\affiliation{
  \institution{Friedrich-Alexander Universit\"{a}t Erlangen-N\"{u}rnberg}
  \city{N\"{u}rnberg}
  \country{Germany}
}

\author{Matthias Stierle}
\email{matthias.stierle@fau.de}
\affiliation{
  \institution{Friedrich-Alexander Universit\"{a}t Erlangen-N\"{u}rnberg}
  \city{N\"{u}rnberg}
  \country{Germany}
  }

\author{Martin Matzner}
\email{martin.matzner@fau.de}
\affiliation{
  \institution{Friedrich-Alexander Universit\"{a}t Erlangen-N\"{u}rnberg}
  \city{N\"{u}rnberg}
  \country{Germany}
}

\author{Stephan Baier}
\email{stephan.baier@fau.de}
\affiliation{
  \institution{Friedrich-Alexander Universit\"{a}t Erlangen-N\"{u}rnberg}
  \city{N\"{u}rnberg}
  \country{Germany}
}

%
\renewcommand{\shortauthors}{Dunzer et al.}
%
\begin{abstract}
Conformance checking is a set of process mining functions that compare process instances with a given process model. It identifies deviations between the process instances' actual behaviour (``as-is'') and its modelled behaviour (``to-be''). Especially in the context of analyzing compliance in organizations, it is currently gaining momentum -- e.g. for auditors. Researchers have proposed a variety of conformance checking techniques that are geared towards certain process model notations or specific applications such as process model evaluation.
This article reviews a set of conformance checking techniques described in 37 scholarly publications.
It classifies the techniques along the dimensions "modelling language", "algorithm type", "quality metric", and "perspective" using a concept matrix so that the techniques can be better accessed by practitioners and researchers. 
The matrix highlights the dimensions where extant research concentrates and where blind spots exist. 
For instance, process miners use declarative process modelling languages often, but applications in conformance checking are rare. Likewise, process mining can investigate process roles or process metrics such as duration, but conformance checking techniques narrow on analyzing control-flow.
Future research may construct techniques that support these neglected approaches to conformance checking.



\end{abstract}

%
%

\ccsdesc[500]{Applied computing~Business process management}
\ccsdesc[300]{Applied computing~Business process modeling}
\ccsdesc[100]{Applied computing~Business intelligence}
\ccsdesc[100]{Information systems~Data mining}

%
\keywords{process mining, conformance checking, modelling languages, business process management, literature review}

%

%

\maketitle

\keywords{
  Process Mining \and Conformance Checking \and Literature Review
}

\section{Introduction}
The Association of Certified Fraud Examiners (ACFE) assembles the \emph{Report to the Nations} once a year. The report's 2018 edition \cite{ACFE.2018} investigates 2,690 cases of professional fraud that were reported between January 2016 and October 2017 in 125 countries. In the average case, an affected company loses about five percent of its annual revenues to occupational frauds and scams. More than a fifth of all reported cases have caused losses of more than one million US dollars. According to the ACFE, either the lack of or the weak implementation of internal fraud control systems (i.e. four-eyes principle) leverages fraudulent behaviour in about half of all reported cases \cite{ACFE.2018}.

All these cases illustrate the need for instruments that support organizations in creating transparency over their internal processes and thereby in identifying irregularities in business process execution and management. Therefore, it is not surprising that conformance checking, one branch of process mining, is gaining momentum in research domain and in practice \cite{aalst2012,greco.2006}. Conformance checking techniques compare the behaviour of process instances recorded in an event log with a process model and analyze deviations of the process behaviour \cite{aalst2012}. Thus, conformance checking measures the deviations of process executions from their normative or descriptive behaviour. As a result, business process managers can either determine whether their processes run as intended or at least as described in a process model. The conformance checking literature is rich on studies linking process models to process data, including the seminal work of \citet{Rozinat2008} and more recent studies as for instance \citet{Burattin2018a}. 
We consider several articles as related works which we divide into three different groups. First, there are several articles that provide literature reviews of process mining projects in a specific field, such as oncology \cite{Kurniati2016}, primary care \cite{Williams2018}, and health care \cite{Rojas2016}. As all these research projects take place in the health care domain and additionally set the broader term 'process mining' as subject to the analysis instead of specializing in conformance checking, the results of our article strongly differ from these previous research projects. Third, \citepos{Rozinat2010} dissertation contains one chapter that deals with conformance measurement using Petri nets. Whereas the author limits the conformance checking overview to Petri nets, we do not filter for one particular process modelling language.
Third, \citet{ElKharbili2008} and \citet{DBLP:books/sp/CarmonaDSW18} explain different foundations of conformance and compliance checking with regard to the process mining domain. On the one hand, \citet{ElKharbili2008} describe different types of compliance checking including \emph{backward compliance checking}. Since we focus on the field of conformance checking in process mining, the present paper follows a different approach. While the authors compare different paradigms of compliance checking, our literature review deals with \emph{backward compliance checking} in by far more detail. \citet{DBLP:books/sp/CarmonaDSW18} recognized the need for a comprehensive overview of this research field in 2018. Therefore, they published the first book dedicated to different concepts of conformance checking. Whereas they focus on the description and explanation of different classes of conformance checking techniques, the present paper provides an overview of the various techniques that were developed and instantiated in the recent years.

Thus, we follow up on \citet{DBLP:books/sp/CarmonaDSW18} by comprising research efforts towards novel conformance checking techniques. Hereby, we outline the techniques' key differences in terms of the supported modelling languages, the concerned perspectives, the employed algorithm types, and the targeted quality metrics.

Therefore, we conduct a systematic literature review to answer the research question:\newline ``\emph{What are the current streams of conformance checking research within the Information Systems discipline and what are their blind spots?}'' 

The present paper provides the outcome of a systematic literature review examining 37 articles that develop distinct conformance checking approaches by answering this research question.

Our main findings consist of a lack of conformance checking techniques for declarative process modelling languages and the need for a greater variety of conformance checking techniques that take further perspectives than just the control-flow of a process into consideration.

Our research objective consists of two types of contributions. First, we contribute to practice by providing an overview of several existing approaches that are relevant for software vendors who implement conformance checking into their products. Second, we provide insights for future researchers in the conformance checking domain. On the one hand, we show existing concepts, whereby we aim to provide a starting point for other researchers. On the other hand, we especially point out identified blind spots of the state-of-the-art literature.

The remainder of the present paper is structured as follows. In the following section, we briefly outline existing concepts of conformance checking to set the theoretical frame for our work. Afterwards, we describe the applied research approach for the systematic literature review. In section \ref{chap:results}, we present the results in the form of a concept-matrix \cite{webster.2002}. Additionally, we describe each of the identified articles' key concepts. Last, a concluding discussion points out the contributions and limitations of the present paper. This section provides an outlook for future research in the domain and concludes the paper.
\section{Conformance Checking Overview}
\emph{Process Mining} is the area within the research field of business process management that deals with the analysis of event data that several information systems generate during the execution of processes. It consists of three distinct sets of functions including discovery, conformance checking, and enhancement \cite{aalst2012}. \emph{Discovery}, the probably most common process mining function, aims to generate as-is process models from event data automatically. \emph{Conformance checking} compares event data that the process execution produces to process models that define their normative or descriptive behavior. Last, \emph{enhancement} includes all functions of process mining that enhance either a process model or an event log. In the remainder, we solely focus on the second set of functions, conformance checking.

We briefly outline existing key concepts of conformance checking regarding its input, its throughput in the form of utilized algorithm types, and its output based on \cite{aalst2012,DBLP:books/sp/CarmonaDSW18}. By examining these basic concepts, we create a theoretical lens for the qualitative analysis of the identified literature in the method section.

As an input, conformance checking methods require an event log and a technical process description. For the scope of the present literature review, we focus on process models with a graphical representation (i.e. Petri nets) as technical process descriptions.
The description of process behavior in line with the specifications of a particular process modelling language creates a process model. Current research distinguishes between two types of process modelling languages. These are, first, the declarative (defining behavior that is not allowed) and, second, the procedural (i.e. defining behavior that is allowed) types of process modelling. The latter type includes popular modelling languages such as Event-driven Process Chains (EPC), Unified Modelling Language (UML) activity diagrams, Business Process Model and Notation (BPMN) and Petri nets \cite{quteprints20105}. Representatives of the declarative modelling languages are Declare, Dynamic Condition Response (DCR) Graphs, Case Management Model and Notation (CMMN) \cite{Jablonski.2016}. 
The event log is the second input for conformance checking. Since the data of events must be relatable to their respective process instance, a clear consensus towards the event log structure exists. Its minimum requirements of content are a unique case identifier, an activity label and a time-stamp \cite[p.98]{vanderAalst:2011:PMD:1983446}. If these content requirements are not fulfilled, conformance checking cannot be performed. Furthermore, event logs may include additional information on further process perspectives, like resources, costs, duration, etc., that conformance checking techniques can utilize for deeper analysis. Hence, the perspectives that a conformance checking technique considers is of high relevance.

After having determined a modelling notation and the scope of the event log (control-flow or multi, i.e. further perspectives), researchers dealing with conformance checking need to choose or develop an algorithm to compare the model to the event log. \citet{DBLP:books/sp/CarmonaDSW18} and \citet{aalst2012} describe two general approaches to such algorithms which are \emph{log replay} algorithms and \emph{trace alignment} algorithms. Log replay algorithms firstly interpret the model and the log and secondly re-run every trace, event by event, on the model. Distinct computing techniques can thereafter determine a conformance metric. One example for such a conformance checking technique is the token-based log replay presented in \cite{Rozinat2008}. Every time the model reaches a dead-end during the execution, before it is allowed to terminate, an additional token is generated to advance the current state of the respective model to the next state. Also, the approach sums up tokens that remained in the model after the log trace has terminated. In the end, the algorithm determines the process conformance based on the sum of all superfluous and generated tokens. Note that this is only one representative example from the variety of existing log-replay algorithms. 

In contrast to log replay algorithms, trace alignments can additionally express deviations and conformance on event level. Figure \ref{fig:example_tracealign} shows the formal notation of trace alignments \cite{aalst2012}.

\begin{figure}[ht]
    \centering
    \begin{equation}
\gamma _1 = 
 \begin{array}{c|c|c|c|c|c}
 a & g & c & f & e & h \\
 \hline
 a & g & c & f & e & h 
 \end{array}
,
\gamma _2 = 
 \begin{array}{c|c|c|c|c|c}
 a & >> & d & b & e & h \\
 \hline
 a & b & d & >> & e & h 
 \end{array}
\end{equation}

\caption{Examples for trace alignments}
\label{fig:example_tracealign}
\end{figure}

 The upper section of an alignment expresses the executed steps in an event log and the lower section shows an aligned trace from the model. Whereas $\gamma _1$ shows one trace alignment that ideally aligns the logged and the modelled trace, $\gamma _2$ contains discrepancies between log and model. The '$>>$' indicates that either the log could not make the same step as in the model or vice versa. By computing such trace alignments, process analysts get various insights in violating log and model traces, as well as the responsible event occurrences. Each asynchronous move in the log or model is related to a cost value to retrieve so-called optimal alignments. Subsequently, the minimum of a cost-function determines optimal alignments by choosing the cheapest alignment of model and log \cite{Adriansyah2013}. 

\begin{figure}[ht]
    \centering
    \includegraphics[width=0.95\columnwidth]{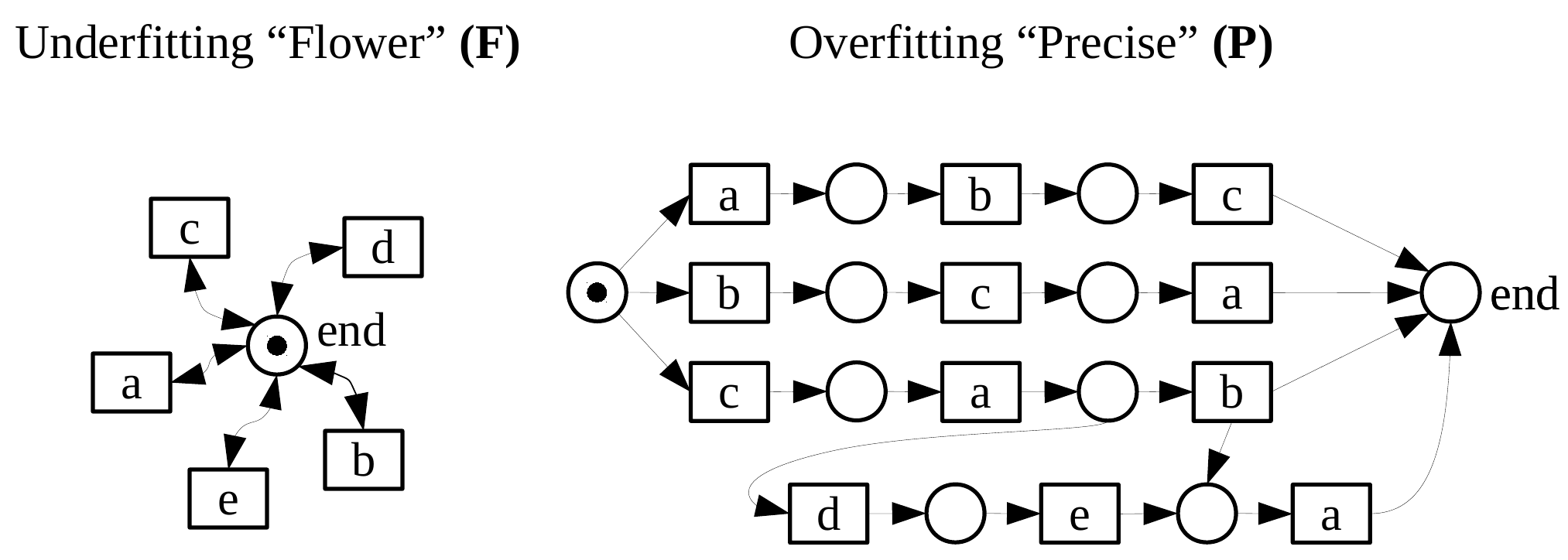}
    \caption{Contrast between generalization and precision \protect\cite{Adriansyah2013}}
    \label{fig:flower}
\end{figure}

Concerning the output of a conformance checking technique, researchers need to decide what metrics they want to use to express conformance. Previous research suggests four quality metrics to measure conformance. \emph{Fitness} expresses the ratio of traces in an event log that a process model can repeat. While a \emph{fitness} of 1.0 indicates that a process model faultlessly reproduces every trace in the event log, a value of 0.0 means that the model cannot repeat any of the real-life cases. \emph{Simplicity} is the quality metric which is often explained with the principle of Occam's razor. It says, the simpler the model is, the better it is. This indicator is useful when the subject of analysis is to find the best model among different process models for the same process. The metric \emph{precision} indicates how well a model represents the behavior seen in a log. \emph{Precision} conformance checking techniques determine superfluous activities and connections to quantify precision. 
In contrast, \emph{generalization} implies whether the process model represents the process paths in a generic way or only the behavior observed in the data. One example for a too general model is the so-called flower Petri net model. Figure \ref{fig:flower} displays the concurrency between the generalization metric represented by the flower model (F) and the over-precise model (P) for the same process. The model F can execute any activity any time, whereby it expresses any process behavior. However, it does not seem to be a good model for a sequential process \cite{VanDerAalst2008,aalst2012,Alizadeh2018}. In contrast, model P depicts every conforming execution path separately and thereby becomes more incomprehensible. As a consequence, researchers must find a balance between generalization and precision for good process models.

Looking at the use of conformance checking in practice, \emph{fitness} is certainly the most used metric as it states how well an observed behavior fits the defined process model, i.e. to evaluate the quality of the executed work. When making statements about the quality of the defined process model process analysts prefer the metrics \emph{precision}, \emph{generalization} and \emph{simplicity} \cite{Aalst2013}.

The analysis of in-, through- and output of the basic conformance checking literature leads to the the following dimensions: (1) \emph{Modelling language}, (2) \emph{Perspective}, (3) \emph{Algorithm type}, and (4) \emph{Quality metric}. These four dimensions categorize all of the conformance checking concepts mentioned above.
\section{Research Method}
We conduct a systematic literature review to provide an overview of the state-of-the-art in conformance checking. The present review grounds on the eight-step guideline delivered by \citet{okoli.2010}. These steps are: (1) purpose of the literature review, (2) protocol and training, (3) searching for the literature, (4) practical screen, (5) quality appraisal, (6) data extraction, (7) synthesis of studies, and (8) writing the review.
Additionally, we take Webster and Watson into consideration as we use a concept matrix as a general framework for the presentation of our results \cite{webster.2002}.

\begin{figure}[htbp]
    \centering
    \includegraphics[width=0.95\columnwidth]{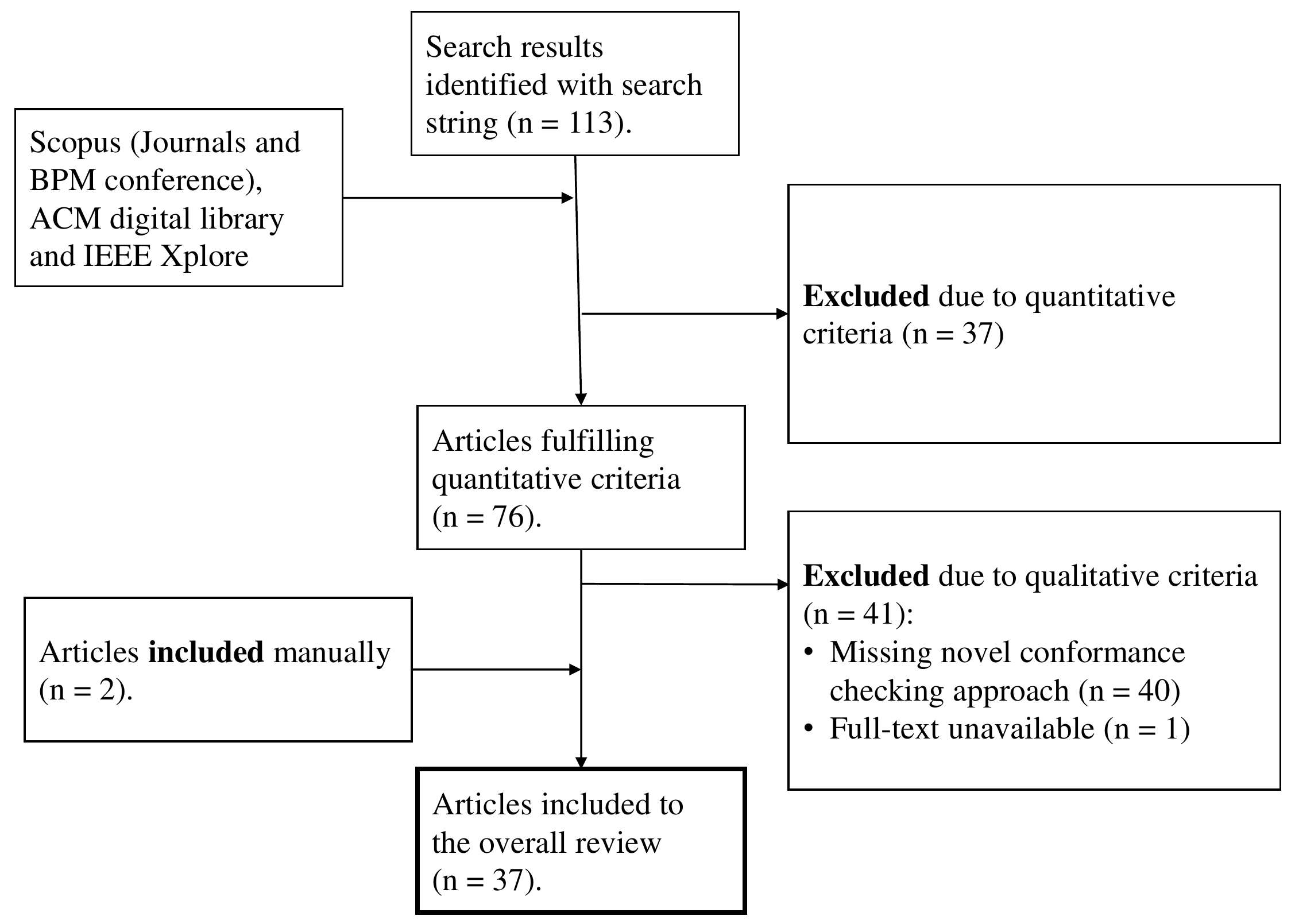}
    \caption{Presentation of literature search process}
    \label{fig:sel_proc}
\end{figure}

The introduction section motivates and explains the \emph{purpose of the literature review}. To conduct the step \emph{protocol and training}, we implemented a spreadsheet that contains all identified articles, a short summary of their core concepts or the justification for their exclusion from the review. 
To identify the first set of literature, we \emph{searched for the literature} using the scientific search engine \emph{Scopus} since it does not focus on a specific research field unlike for example primary databases. However, to minimize the chances of leaving out essential articles we also included the databases \emph{ACM Digital Library} as well as \emph{IEEE Xplore} in our search.
We used the keywords \emph{conformance checking}, \emph{process} and \emph{mining} as terms that need to be contained in the title of the article, abstract or keywords.
Furthermore, we include only articles that were already published in a peer-reviewed journal or the topic related conference \emph{BPM} in the analysis. Thus, all documents that still are in-press are also dismissed. Furthermore, we limit the publication language to English. This literature search identifies 113 articles in total. 

Since the present paper displays current streams of the conformance checking literature we focus on the most recent articles as well as articles that are of significant influence to the current streams. To meet these requirements, we assume that articles that have been published three years ago and do not have a significant \emph{Cited by} number are of rather less relevance to the present research. Note that, we do not exclude author self-references to limit the article list. Thus, the following quantitative criteria were applied sequentially during the step \emph{practical screening} to limit the results to the relevant articles for the present systematic literature review:

\begin{enumerate}
\item Number of \emph{Cited by} is higher than ten and \emph{Publication year} is before '2017'.
\item Number of \emph{Cited by} is higher than two and \emph{Publication year} is '2017'
\item \emph{Publication year} is '2018' or '2019'
\end{enumerate}

If an article does not match one of the mentioned criteria it was excluded from the set of relevant literature. For the evaluation of bibliometrics, we used Scopus' \emph{Cited by} number. All in all, 37 articles were eliminated during this step of the analysis. During the conduction of the \emph{quality appraisal}, we skimmed the abstracts, results, and discussions of the remaining 76 articles to decide whether the articles closely relate to conformance checking. Furthermore, the relevant articles are required to deliver an own approach to perform conformance checking. As a result, we identified 40 articles that do not primarily deal with conformance checking or do not deliver a novel conformance checking approach. One article was not available as full-text to us. While examining the articles, we searched for eventually missing literature within a forward as well as backward search. In this process, we added two articles to our set of literature that are not part of the results of the literature search but fit the requirements of the present review. In the step \emph{data extraction}, the remaining 37 articles are clustered regarding their relation to the concept dimensions elaborated in the previous section. The resulting concept matrix shows which concepts have and which have not been researched thoroughly \cite{webster.2002}. Thus, we consider the concept matrix as our \emph{synthesis of studies}. In the following section, we describe every dimensions and their related concepts by \emph{writing the review}.

\section{Results}
\label{chap:results}
This section presents the key findings of the systematic literature review. We present the concept matrix in Table \ref{fig:conceptmatrix} by describing current research streams for each of the dimensions. Note that, during the examination of the articles we encountered two additional algorithm types which is why we added the column \emph{others} to this dimension. Attentive readers might wonder how some articles, e.g. \cite{Kalenkova2017,Burattin2016,JagadeeshChandraBose2012,Alizadeh2018,DeLeoni2015}, check conformance without assessing a particular quality metric. This is a consequence of approaches that convert process models from one language into another to perform conformance checking. Likewise, some authors declare a conformance value that does not exactly fit the definition of one of the quality metrics. Furthermore, if there is no model marked in the modelling language dimension, the authors of the examined article claim their approach to be independent of the modelling language.

\subsection{Modelling language}
Concluding from the first dimension \emph{modelling language} in the concept matrix presented in \autoref{fig:conceptmatrix}, Petri nets embody by far the most commonly used process modelling language in conformance checking literature with 28 occurrences. A Petri net is a visually presentable graph expressing execution logic such as restrictions, control-flow, and concurrency. Due to their mathematical background, their capability to capture concurrent behavior, and their function as state charts, Petri nets have been thoroughly developed as a formal process modelling language. \citet{Burattin2018a}, for instance, use Petri nets as a representative for all procedural modelling languages in their conformance checking framework. Even though most of the articles that use Petri nets mainly evaluate the control-flow of a process, there are a few approaches concerning multiple perspectives. Whereas, \citet{Mannhardt2016}, and \citet{DeLeoni2013} utilize so-called \emph{data Petri nets} to enhance Petri nets by means to evaluate data-constraints, \citet{Alizadeh2018} employ a so-called \emph{"Create Read Update Delete" (CRUD) matrix}. The CRUD matrix relates data objects to process logic (i.e. a Petri net places or transitions), so that an activity occurs only if the related data allows or enforces it. The data Petri nets attach data-constraints wherever necessary to places and transitions of a Petri net. Subsequently, the algorithms skim the event data related to the current state of the process execution \cite{Mannhardt2016,DeLeoni2013}. Thus, enhanced Petri nets accomplish complete conformance checking that evaluates the control-flow and the data-perspective.\newline
In the subset of the literature, three articles use BPMN as the technical specification for their input model. Although BPMN is quiet popular in the corporate environment, only one of the three approaches conducts conformance checking with native BPMN \cite{leoniBpmn2012}. The authors state that their approach is independent from the actual modelling language since it takes advantage of extended casual nets. As a result, the authors argue that process logic is generally expressible in extended causal nets. While \cite{leoniBpmn2012} use native BPMN process models to describe their conformance checking approach, \cite{Kalenkova2017} and \cite{Garcia-Banuelos2018} utilize a different technique to perform conformance checking using BPMN process models. They develop an artifact to convert BPMN process models into Petri nets so that they can perform conformance checking with the resulting Petri net. \newline
Last, a few articles do not focus on process modelling languages but instead abstractly describe conformance concepts using, for instance, so-called work-flow descriptions \cite{JagadeeshChandraBose2012,JagadeeshChandraBose2010,Munoz-Gama2014}. \newline
The remaining articles use Declare process models as an input. Two of the identified articles conduct multi-perspective conformance analysis by implementing business rules and data constraints \cite{Burattin2016,Borrego2014}. Declare is presently the only graphical declarative language that has been employed in conformance checking research. 

\subsection{Perspective}
The \emph{perspective} dimension shows that six articles develop multi-perspective and 31 articles control-flow perspective approaches. As further data and perspectives, the multi-perspective articles mostly used time, roles, and contextual data such as costs \cite{Alizadeh2018,Borrego2014,Burattin2016,Mannhardt2016,leoniBpmn2012}. \newline
While performing conformance checking of process models based on constraints, these process-flow constraints can be enhanced by business data constraints to project the data-flow \cite{Borrego2014}. \citet{Alizadeh2018} provide means to conduct extended process auditing, whereby they combine the process and data perspectives to perform conformance checking. \Citet{DeLeoni2013} focus on the additional perspectives of data, time and resource, whereas one of their previous articles takes only data and resources into consideration \cite{leoniBpmn2012}. One article addresses a combination of the process, time, and data perspectives in combination with declarative business rule checking \cite{Burattin2016}. Another notable aspect in this particular dimension is that most of the approaches that use declarative approaches can include more perspectives than the control-flow. Although a few authors state that conformance checking techniques must be able to detect deviating behavior concerning the misuse of organizational resources \cite{leoniBpmn2012}, the number of existing articles itself indicates that the control-flow of processes is of higher relevance for the process mining domains. 

\begin{table}																			
\caption{Concept matrix using dimensions following \cite{webster.2002}}																			
\label{fig:conceptmatrix}																			
\centering\settowidth\rotheadsize{Trace alignment/}																			
  \renewcommand\cellalign{cc}																			
  \renewcommand\arraystretch{1.25}																			
\resizebox{\columnwidth}{!}{%
\begin{tabular}{|p{3cm}|>{\centering\arraybackslash}p{0.75cm}|>{\centering\arraybackslash}p{0.75cm}|>{\centering\arraybackslash}p{0.75cm}|>{\centering\arraybackslash}p{0.8cm}|>{\centering\arraybackslash}p{0.8cm}|>{\centering\arraybackslash}p{0.55cm}|>{\centering\arraybackslash}p{0.55cm}|>{\centering\arraybackslash}p{0.55cm}|>{\centering\arraybackslash}p{0.5cm}|>{\centering\arraybackslash}p{0.5cm}|>{\centering\arraybackslash}p{0.5cm}|>{\centering\arraybackslash}p{0.5cm}|}																			
\toprule\noalign{\vskip-1pt}\hline																			
	& \multicolumn{3}{|c|}{\textbf{Modelling language}}			& \multicolumn{2}{|c|}{\textbf{Perspective}}		& \multicolumn{3}{|c|}{\textbf{Algorithm type}}			& \multicolumn{4}{|c|}{\textbf{Quality metric}}						\\ \hline				
\diagbox[height=1.25\rotheadsize,innerwidth=3cm]{\raisebox{3ex}{\textbf{Articles}}}{\raisebox{-4ex}{\textbf{Concept}}} 	& \rotcell{Declare}	& \rotcell{BPMN}	& \rotcell{Petri nets}	& \rotcell{Control-flow}	& \rotcell{Multi}	& \rotcell{Log replay}	& \rotcell{Trace alignment}	& \rotcell{Others}	& \rotcell{Fitness}	& \rotcell{Simplicity}	& \rotcell{Precision}	& \rotcell{Generalization}			\\ \hline				
\cite{Rozinat2008}	 & 	 & 	& \textbullet	& \textbullet	 & 	& \textbullet	 & 	 & 	& \textbullet	 & 	& \textbullet	& \textbullet			\\ \hline				
\cite{VanDerAalst2008}	 & 	 & 	& \textbullet	& \textbullet	 & 	& \textbullet	 & 	 & 	& \textbullet	 & 	& \textbullet	& \textbullet			\\ \hline				
\cite{JagadeeshChandraBose2010}	 & 	 & 	 & 	& \textbullet	 & 	 & 	& \textbullet	 & 	 & 	 & 	 & 	 & 			\\ \hline				
\cite{Adriansyah2011}	 & 	 & 	& \textbullet	& \textbullet	 & 	& \textbullet	 & 	 & 	& \textbullet	 & 	 & 	 & 			\\ \hline				
\cite{JagadeeshChandraBose2012}	 & 	 & 	 & 	& \textbullet	 & 	 & 	& \textbullet	 & 	 & 	 & 	 & 	 & 			\\ \hline				
\cite{DeLeoni2012}	& \textbullet	 & 	 & 	& \textbullet	 & 	 & 	& \textbullet	 & 	& \textbullet	 & 	 & 	 & 			\\ \hline				
\cite{leoniBpmn2012}	 & 	& \textbullet	 & 	& \textbullet	& \textbullet	 & 	& \textbullet	 & 	& \textbullet	 & 	 & 	 & 			\\ \hline				
\cite{VanDerAalst2013}	 & 	 & 	& \textbullet	& \textbullet	 & 	 & 	& \textbullet	 & 	& \textbullet	 & 	 & 	 & 			\\ \hline				
\cite{Adriansyah2013}	 & 	 & 	& \textbullet	& \textbullet	 & 	 & 	& \textbullet	 & 	 & 	 & 	& \textbullet	 & 			\\ \hline				
\cite{Aalst2013}	 & 	 & 	& \textbullet	& \textbullet	 & 	& \textbullet	 & 	 & 	& \textbullet	& \textbullet	& \textbullet	& \textbullet			\\ \hline				
\cite{DeLeoni2013}	 & 	 & 	& \textbullet	& \textbullet	& \textbullet	 & 	& \textbullet	 & 	& \textbullet	 & 	 & 	 & 			\\ \hline				
\cite{Kirchner2013}	 & 	 & 	& \textbullet	& \textbullet	 & 	 & 	& \textbullet	 & 	& \textbullet	 & 	 & 	 & 			\\ \hline				
\cite{Adriansyah2013a}	 & 	 & 	& \textbullet	& \textbullet	 & 	 & 	& \textbullet	 & 	& \textbullet	 & 	& \textbullet	 & 			\\ \hline				
\cite{munozgama2013}	 & 	 & 	 & 	& \textbullet	 & 	& \textbullet	 & 	 & 	& \textbullet	 & 	 & 	 & 			\\ \hline				
\cite{Munoz-Gama2014}	 & 	 & 	& \textbullet	& \textbullet	 & 	 & 	& \textbullet	 & 	& \textbullet	 & 	 & 	 & 			\\ \hline				
\cite{Adriansyah2014}	 & 	 & 	& \textbullet	& \textbullet	 & 	 & 	& \textbullet	 & 	 & 	 & 	& \textbullet	 & 			\\ \hline				
\cite{VandenBroucke2014}	 & 	 & 	& \textbullet	& \textbullet	 & 	 & 	 & 	& \textbullet	 & 	 & 	& \textbullet	& \textbullet			\\ \hline				
\cite{VanDerAalst2014}	 & 	 & 	& \textbullet	& \textbullet	 & 	& \textbullet	 & 	 & 	& \textbullet	 & 	 & 	 & 			\\ \hline				
\cite{Borrego2014}	& \textbullet	 & 	 & 	& \textbullet	& \textbullet	 & 	 & 	& \textbullet	 & 	 & 	 & 	 & 			\\ \hline				
\cite{Fahland2015}	 & 	 & 	& \textbullet	& \textbullet	 & 	 & 	& \textbullet	 & 	& \textbullet	& \textbullet	& \textbullet	& \textbullet			\\ \hline				
\cite{DeLeoni2015}	& \textbullet	 & 	 & 	& \textbullet	 & 	 & 	& \textbullet	 & 	& \textbullet	 & 	& \textbullet	& \textbullet			\\ \hline				
\cite{Lu2015}	 & 	 & 	& \textbullet	& \textbullet	 & 	 & 	& \textbullet	 & 	 & 	 & 	 & 	 & 			\\ \hline				
\cite{Verbeek2015}	 & 	 & 	& \textbullet	& \textbullet	 & 	 & 	& \textbullet	 & 	 & 	 & 	 & 	 & 			\\ \hline				
\cite{Mannhardt2016}	 & 	 & 	& \textbullet	& \textbullet	& \textbullet	 & 	& \textbullet	 & 	& \textbullet	 & 	 & 	 & 			\\ \hline				
\cite{Burattin2016}	& \textbullet	 & 	 & 	& \textbullet	& \textbullet	& \textbullet	 & 	 & 	 & 	 & 	 & 	 & 			\\ \hline				
\cite{Kalenkova2017}	 & 	& \textbullet	 & 	& \textbullet	 & 	 & 	& \textbullet	 & 	 & 	 & 	 & 	 & 			\\ \hline				
\cite{Leoni2017}	 & 	 & 	& \textbullet	& \textbullet	 & 	 & 	& \textbullet	 & 	 & 	 & 	 & 	 & 			\\ \hline				
\cite{Song2017}	 & 	 & 	& \textbullet	& \textbullet	 & 	 & 	& \textbullet	 & 	 & 	 & 	 & 	 & 			\\ \hline				
\cite{Garcia-Banuelos2018}	 & 	& \textbullet	 & 	& \textbullet	 & 	 & 	& \textbullet	 & 	 & 	 & 	 & 	 & 			\\ \hline				
\cite{Leemans2018}	 & 	 & 	& \textbullet	& \textbullet	 & 	& \textbullet	 & 	 & 	& \textbullet	 & 	& \textbullet	& \textbullet			\\ \hline				
\cite{Alizadeh2018}	 & 	 & 	& \textbullet	& \textbullet	& \textbullet	 & 	& \textbullet	 & 	 & 	 & 	 & 	 & 			\\ \hline				
\cite{Koorneef2018}	 & 	 & 	& \textbullet	& \textbullet	 & 	 & 	& \textbullet	 & 	& \textbullet	 & 	 & 	 & 			\\ \hline				
\cite{Lee2018}	 & 	 & 	& \textbullet	& \textbullet	 & 	 & 	& \textbullet	 & 	& \textbullet	 & 	 & 	 & 			\\ \hline				
\cite{Bloemen2018}	 & 	 & 	& \textbullet	& \textbullet	 & 	 & 	& \textbullet	 & 	& \textbullet	 & 	 & 	 & 			\\ \hline				
\cite{Burattin2018}	 & 	 & 	& \textbullet	& \textbullet	 & 	 & 	 & 	& \textbullet	& \textbullet	 & 	 & 	 & 			\\ \hline				
\cite{Dongen2018}	 & 	 & 	& \textbullet	& \textbullet	 & 	 & 	& \textbullet	 & 	 & 	 & 	 & 	 & 			\\ \hline				
\cite{Burattin2018a}	 & 	 & 	& \textbullet	& \textbullet	 & 	 & 	& \textbullet	 & 	& \textbullet	 & 	& \textbullet	 & 			\\ \hline				
\toprule\hline					
\end{tabular}				
}																\end{table}

\subsection{Algorithm type}
In this section, we elaborate our findings regarding the algorithm types that the approaches apply to perform conformance checking. Notably, we encountered two additional \emph{algorithm types} during the examination of the articles. Therefore, we expand the two initial concepts of trace alignment and log replay by constraint-based conformance checking and artificial negative event precision checking in the concept matrix column "others". \newline
\Citet{VandenBroucke2014} design the artificial negative event algorithm aiming to improve models by checking their precision and generalization in a novel way. Here, they define negative events as events that should not occur at a specific time or state of a process execution. When the trace $t\{a , b, c\}$ is a successfully executing process trace in the event log $L$, then the $t'$, with induced artificial negative events for $t$, is $t'\{(b^-, c^-), a, (a^-, c^-), b, (a^-, b^-), c\}$. Here, the superscript '$-$' indicate negative events and '$()$' artificially generated events. To determine the conformance metric precision, every traces in the event log $L$ is replayed including the identified negative events from trace $ t' $. If a negative event never occurs, the process path including the negative event can be removed from the model. Otherwise, if it occurs as often as previously defined, a new process path should be added to the model. 

\citet{Borrego2014} and \citet{Burattin2018a} perform constraint-based conformance checking. Thus, they translate process logic into rules and constraints which can thence check conformance by evaluating the fulfillment of each rule in every trace. Furthermore, the rules may contain data related constraints which leverage multi-perspective conformance checking \cite{Borrego2014}. The other constraint-based approach employs a behavioral model approach for conformance checking \cite{Burattin2018a}. For this technique the authors assume that a process analyst can derive more advanced behavioral patterns from an existing process model. These behavioral patterns may express for example that activity $\alpha$ must occur before $\beta$. After having derived all behavioral patterns from the model, these are evaluated with all process traces in an event log.\newline
All in all, seven articles in the present set of literature use the log replay as an algorithmic base to determine the conformance of a log or a model. In order to perform conformance checking with a log-replay algorithm, an algorithm designer needs to implement a model interpreter beforehand. The model interpreter extracts all the rules and allowed states as well as transitions from a process model. After that, the algorithm re-runs and re-evaluates every trace from the event log, whereby the interpreted model specifications imply whether the replay fails or succeeds for each trace. There are several ways to measure conformance using log-replay algorithms. For instance, the token-based log replay (see section 2) \cite{VanDerAalst2008,VanDerAalst2013,Rozinat2008}. In contrast to the token-based log-replay algorithm, creating a deterministic finite automaton from the input model first and thence replay the event log on the allowed traces of the automaton does not make use of tokens itself \cite{Leemans2018}.
\citet{Burattin2016} design another approach to conduct conformance checking with declarative process models based on log replay. They implement a model interpreter to single out linear temporal logic constraints from the input Declare model and then the approach checks every trace for the fulfillment or violation concerning these constraints. One article addresses the problem of large event logs in process mining. By using the single entry single exit approach to divide a process model into sub-models, they replay each log part in its corresponding single entry single exit sub-model \cite{munozgama2013}. Last, there is the technique of cost-based replay of the model \cite{Adriansyah2011}. This method explicitly applies costs to the arcs of a Petri net and to inserted or skipped activities. Afterwards, they calculate conformance ratios based on the outcome costs of the algorithm.

The major part of the identified literature employs trace alignment algorithms to compute the conformance of processes. Thus, trace alignments can be considered as the current standard of conformance checking techniques \cite{Lee2018}. Trace alignment algorithms mostly utilize procedural process modelling languages. Two articles by the same main author promote a quite similar alignment framework. \Citet{JagadeeshChandraBose2012} and \citet{JagadeeshChandraBose2010} execute the following steps to align multiple traces: pre-processing of event logs, computing scoring matrices, building a guide tree for multiple trace alignment, estimating the alignment quality, pruning and realignment, and an interactive visualization. Some of the research projects expanded the basic approaches with additional features to compute optimal alignments. Especially, since the A* algorithm returns multiple complete alignments for each log and model trace, a means to retrieve the optimal alignment is required \cite{Adriansyah2011}. Therefore, cost functions are defined and evaluated for each alignment to get the optimal trace alignment based on the optimization of a cost function \cite{Adriansyah2013,Adriansyah2014,Leoni2017}. These optimization functions rely on several types of input. These consist of distance  \cite{Adriansyah2013,Adriansyah2014}, manually customizable costs for specific process violations \cite{DeLeoni2013}, legal moves \cite{leoniBpmn2012,Leoni2017}, probabilities \cite{Koorneef2018}, synchronous moves \cite{Bloemen2018}, region theory and state similarity \cite{Burattin2018}. Furthermore, two approaches implement cost-functions to create alignments which are customized to event monitoring points in a hospital setting to compute optimal alignments \cite{Kirchner2013,Adriansyah2011}. Besides a cost function to compute an optimal alignment, another approach inserts the planning domain definition language to formulate computing alignment as a so-called planning problem \cite{Leoni2017}. Subsequently, off-the-shelf automated planners can resolve optimal alignments based on the determined planning problem. This approach bears the advantage of memory-efficiency. Another method promoting memory-efficient alignment computing also derives from the automated planning domain \cite{Adriansyah2013a}. Whereas these two concepts focus on efficient usage of memory, \citet{Dongen2018} elaborates an algorithm based on estimated heuristics to create a CPU-efficient and faster solution to compute trace alignments.
Researchers have advanced the control-flow alignment approach to be capable of handling multi-perspective process analysis \cite{Mannhardt2016,DeLeoni2013,leoniBpmn2012,Alizadeh2018}.  
\citet{Mannhardt2016} evolve basic alignments into balanced multi-perspective alignments that align resource and context-data between an event log and a model. Another approach implements an integer linear programming approach by adding the dimensions of time, data and resources to alignment cost functions \cite{DeLeoni2013}. Furthermore, an approach to create multi-perspective alignments with BPMN process models uses causal nets enhanced with data \cite{leoniBpmn2012}. One technique enhances trace alignments to inter-level alignments that relates process steps to log steps and to respective data entries in a CRUD matrix to enable multi-perspective conformance checking \cite{Alizadeh2018}. Moreover, a method converts a BPMN process model and its respective event log into primary event structures \cite{Garcia-Banuelos2018}. Consequently, both structures can be merged to generate an alignment between the model and the log.
\citet{Fahland2015} focus on model repair, whereby they use an alignment approach to determine missing or superfluous activities and connections. A variety of trace alignment approaches address the issue of decomposing complex conformance checking scenarios \cite{VanDerAalst2013,Munoz-Gama2014,Verbeek2015,Song2017}. One Petri net decomposition splits the overall process model into sub-nets \cite{VanDerAalst2013}. As a result, these sub-nets can be aligned with certain parts of traces in an event log. The single entry single exit concept adds to the conformance checking approaches dealing with large event logs \cite{Munoz-Gama2014}. If a Petri net contains sequences that start at a single place (entry), then run through different process paths, and reach the same end node in any case, this sequence can be extracted and evaluated individually as it is considered as a valid sub-net. Additionally, a generic divide-and-conquer framework supports the decomposition of process models \cite{Verbeek2015}. \citet{Song2017} propose alignments based on heuristics rather than the A* algorithm and the divide-and-conquer strategy. From this point on they utilize the alignments to design a model repair method for industrial scale processes. 
Even though the lion's share of trace alignment research is dedicated to procedural modelling languages, there are artifacts to create alignments using the declarative process modelling languages \cite{DeLeoni2012,DeLeoni2015}.
All the above-mentioned approaches rely on clean and correctly ordered event logs. To address the issue of only partially ordered event logs, \citet{Lu2015} designed a trace alignment approach that calculates partial alignments by setting up dependencies for each place instead of strictly focusing on a place. The last concept of creating trace alignments converts BPMN process models into Petri nets, so that the optimal trace alignments can be computed for the resulting Petri net instead of the input BPMN model \cite{Kalenkova2017}.

\subsection{Quality metric}
In the present systematic literature review, four \emph{Quality metrics} to measure conformance for event logs and models are identified. These metrics are fitness, simplicity, precision, and generalization.

Given that the main objective of conformance checking is to identify whether the execution of the process (i.e. the data) is conforming with the specified process (i.e. the model), it is not surprising that fitness is the quality measure most frequently used throughout the examined articles. There are several ways to measure fitness. First, by using the log replay algorithm, the number of re-playable log traces can be ascertained. The division of the re-playable traces by the number of all traces can indicate a fitness value \cite{Fahland2015,VanDerAalst2008,munozgama2013,Leemans2018}. Second, there is the token-based log-replay method to calculate the fitness (see conformance checking overview section). \Citet{Adriansyah2011} present such an event-based fitness measure that counts each non-conforming event and divides them by the total number of events. Third, trace alignments utilize different kinds of cost functions to compute a fitness metric \cite{JagadeeshChandraBose2010,DeLeoni2012,DeLeoni2013,Kirchner2013,Adriansyah2013a,VanDerAalst2014,Munoz-Gama2014,Fahland2015,DeLeoni2015,Koorneef2018,Bloemen2018}. 
Two declarative ones approach the fitness calculation regarding previously determined business rules \cite{Borrego2014,Burattin2018a}. The approach records every trace whereby the beforehand defined business rules indicate whether a trace is sound or not at the time of the execution \cite{Borrego2014}. By dividing the overall number of recorded traces by the number of violating traces, a process analyst can determine a conformance ratio already during the run-time of the respective process. Additionally, these rules may refer to event data, such as costs, resources etc., whereby the approach calculates multi-perspective fitness values for processes \cite{Burattin2018a}.
Because trace alignments compute fitness values on the event-level instead of the trace level, they are considered to be an accurate tool to determine fine-grained conformance information \cite{Mannhardt2016}. Although this method of fitness calculations is most capable, some drawbacks come with it. Finding an optimal alignment between large event logs and a model is considered as an NP-hard problem since an increasing number of iterative routes in a process model increases the number of potential alignment paths to a considerable extent. \citet{Lee2018} tackle this problem by proposing a decomposed fitness measure that shows significant improvements in performance while getting very close to the overall fitness measure based on the optimal alignment.

Another metric to assess model quality relates to the principle of Occam's razor and is called simplicity. When concerning simplicity, a process analyst assumes that the simplest model that expresses the behavior of a process model is always the best model. However, simplicity is hard to quantify and often refers to the number of nodes, and transitions in a process model as quantifiers \cite{Garcia-Banuelos2018}. A distinct approach defines simplicity as the number of sub-models added in the log compared to the existing process model \cite{Fahland2015}.

The two remaining metrics, precision and generalization, are contradicting themselves. The model should be as general -- that is, the model should not overfit the log -- and as precise -- that is, the model should not underfit the log -- as possible \cite{DeLeoni2015}. The majority of the identified articles utilize an alignment-based precision metric that is the result of dividing the available actions in the model by the executed actions \cite{Adriansyah2013,Adriansyah2014,Adriansyah2013a,VanDerAalst2013}. Similarly, \citet{DeLeoni2015} calculate precision for their declarative approach by dividing all executed activity sequences by the number of possible activity sequences. Furthermore, they use a probabilistic approach based on the alignment automaton of the Declare process model to estimate the probability of new events occurring that are not reflected in the event log \cite{DeLeoni2015}.
One technique leverages the calculation of precision and generalization by using model-to-model comparison \cite{Leemans2018}. The authors compare the model generated from the data with the provided process model and divide them into smaller sub-models in order to reduce the complexity. An especially different method to measure precision and generalization implements the artificial negative event approach \cite{VandenBroucke2014}. Here, the log is replayed with artificial negative events. By utilizing the number of replayed positive (allowed occurrence) and weighted negative (erroneous occurrence) events, the precision and generalization metric is calculated. This approach leverages process analysts to determine the exact point of the model that allows or restricts too much behavior. 

\section{Discussion of the findings and future research directions}

The present literature review examines the state of the existing techniques and reveals aspects of conformance checking that have not been examined as thoroughly as others. For that, we analyzed 37 articles to outline differences regarding \emph{Modelling language}, \emph{Perspective}, \emph{Algorithm type} and \emph{Quality metric}. In the following subsections, we discuss the findings of our review, state its limitations and conclude with an outlook for future activities building on our work.

\subsection{Findings}

Our literature review reveals that most articles examine the quality metric fitness regarding the quality metrics computed by conformance checking techniques. After all, this is not a big surprise given the fact that especially practitioners want to assess the quality of process execution rather than to assess the quality of the defined process model. In addition to the various concepts to calculate fitness, many researchers improve alignment based approaches in terms of their cutstomizability regarding their cost-functions. 
Besides fitness, a few approaches deal with precision and generalization of process models. Simplicity, however, is rarely discussed in detail and there is no general quantitative approach to calculate it. After all, simplicity seems to result from the trade-off between precision and generalization which is hard to quantify.
Regarding the algorithm types, techniques from the early days of conformance checking employ log replay algorithms. Today, almost every article builds on the concept of trace alignments. While trace alignments offer powerful capabilities, they come with the disadvantage of the complexity related to the pre-computing of model traces. Some articles propose solutions for more efficient ways to calculate aligned traces. Nevertheless, this area requires more research.

Regarding the process modelling languages, our research reveals a strong focus on procedural models specifically on Petri nets. Although research about procedural models is widely spread, regarding perspectives, algorithm types, and quality metrics, we find it noticeable that the variety of existing procedural modelling techniques, like event process chains, subject-oriented business process management or even BPMN, is not well covered in the state-of-the-art literature. Moreover, only four articles use the only representative of declarative process modelling languages, Declare. This fact provides evidence for a gap between the research efforts towards procedural and declarative process modelling. However, more and more research in the field of business process management is dedicated to declarative languages because they promise potential when dealing with complex and unstructured processes \cite{DeLeoni2012,Borrego2014}. While Declare is undoubtedly a prominent representative of the declarative languages, it has some drawbacks like slow adaption by users due to its complexity \cite{Haisjackl2016}. Therefore, we argue to consider further declarative languages for the development of new conformance checking techniques.

Last but not least, more than 80\% of all reviewed articles approach conformance checking by exclusively examining the control-flow of processes omitting further perspectives. It is astonishing to a certain extent that only six articles develop conformance checking techniques taking a multi-perspective view on processes and thereby being able to check further constraints. Therefore, we call for multi-perspective conformance checking techniques that allow a holistic conformance analysis of the process. A challenge for such techniques will certainly be posed by the increasing complexity and hence, the need for efficient techniques in order to maintain performance \cite[p.262]{DBLP:books/sp/CarmonaDSW18}. 

\subsection{Contributions}
We contribute to academia by providing an overview of the state-of-the-art to researchers who work in the domain of conformance checking. Such an overview helps to steer future research by outlining research gaps and subsequently displaying research needs for new techniques. Furthermore, the concept matrix in the results section shows exactly, which topics and concepts are not as well researched as others. Thus, our review may also function as a starting point to address new concepts in the research area. 

Besides our contribution to the research domain, we provide a well-structured overview of recent relevant conformance checking approaches including a categorization of each of the identified techniques. This is particularly helpful for software vendors aiming at including conformance checking into their products. It might give business process management in corporate environments an overview of all the dimensions they may employ as well. Thus, we believe that our artifact supports managers' decisions regarding the distinct dimensions when they want to implement conformance checking for their processes. 

\subsection{Limitations}
Despite our best efforts to undertake a comprehensive search and review of the articles, we point out a few limitations to our review. Some of the articles were excluded due to the language barrier as they were not available in English which may have resulted in omitting important papers. Another minor limitation of our review might be the usage of Scopus for the evaluation of bibliometrics since the articles may have different numbers of citations on different search engines. However, we chose Scopus since it covers several research fields and numerous peer-reviewed journals and conference papers and does not customize search results based on preferences. Although the application of quantitative criteria in the literature selection might lead to missing out a "late-blooming" article, we believe that we captured the general picture of conformance checking approaches that use a process model and an event log as input. One of the articles meeting the quantitative criteria was not available as a full-text version to us. 
Despite these limitations, the review provided important insights into existing conformance checking techniques, their differences and blind-spots.

\subsection{Outlook}
We call for further research in the area of conformance checking based on the findings of our systematic literature review. In the following, we want to present how future researchers may extend our work to provide further insight into the field of conformance checking. First, we see potential in a more detailed review of articles working with trace alignments to identify differences in terms of their implementations. As trace alignments seemingly present the most suitable algorithm type for conformance checking, this would help to steer future research in the right direction. Especially, differences regarding their applied optimization-functions to compute optimal alignments might be important to practice and future research. Second, an evaluation of existing conformance checking techniques with real-life event logs should be conducted in order to assess the potential of conformance checking for organizations. Thus, the validation of some of the current conformance checking results lack empirical data, although we found several articles about conformance checking used in hospitals. Still, only few articles about the real-life application in distinct companies exist. Insights about the feasibility of conformance checking techniques in practice and potential benefits would help to identify requirements for new techniques. Last, we encountered two articles in our keyword search concerned with constraint- or rule-based conformance checking. Since we want to streamline strictly the "conformance checking" keyword which is solely related to the set of functions delivered in process mining. Thus, in future research, it might be of good value to provide an in-depth analysis of non-graphical process definitions, such as business rules etc., as well. Furthermore, related research fields such as fraud and anomaly detection, and compliance checking could be skimmed for articles that are of relevance to the conformance checking research domain.

\subsection{Conclusion}
With organizations looking for ways to efficiently and effectively manage their processes, and to avoid fraud and non-compliance, conformance checking is gaining momentum in practice. In the research domain, the number of publications is rising year by year. Thus, we wanted to identify trends and shortcomings of the conformance checking research of recent years to streamline the academic work. To address this objective, we conducted a systematic literature review of existing conformance checking techniques. We classified the 37 identified articles regarding the dimensions of \emph{Modelling language}, \emph{Perspective}, \emph{Algorithm type} and \emph{Quality metric}. The classification of the articles is displayed in a concept-matrix that relates distinct concepts of the particular dimension to the article. First, the identified \emph{Modelling languages} consist of BPMN, Declare and Petri nets. Second, we classified the concepts belonging to the dimension \emph{Perspective} by the control-flow perspective and multi-perspective approaches. Third, we encountered four different \emph{Algorithm types}, whereby we summarized the approaches of rule-based conformance checking and artificial negative events as \emph{Others}. The remaining two \emph{Algorithm types} are the log-replay type and the most popular approach, the so-called trace alignments. Last, we found four distinct \emph{Quality metrics}, fitness, simplicity, precision, and generalization. 
The systematic literature review presented in this research reveals that there is a lack of techniques based on declarative modelling languages. Furthermore, we observe that conformance checking techniques currently focus on the analysis of the process control-flow. To conclude the present paper, we want to encourage future researchers to address multi-perspective techniques that take account of further perspectives like time or roles so that organizations can analyze their complex business processes from various angles.

\bibliography{sample-base}
\bibliographystyle{ACM-Reference-Format.bst}
\end{document}